\documentclass[twocolumn,english,aps,prb]{revtex4}
\usepackage[T1]{fontenc}
\usepackage[latin9]{inputenc}
\usepackage{graphicx}
\usepackage{amssymb}

\makeatletter
\makeatletter


\usepackage{babel}

\makeatother

\usepackage{babel}

\makeatother

\usepackage{babel}

\begin{document}

\title{Gd$^{3+}$ rattling triggered by a {}``weak'' M-I transition at
140-160 K in the Ce$_{1-x}$Gd$_{x}$Fe$_{4}$P$_{12}$ $x\approx0.001$
skutterudite compounds: an ESR study.}

\author{F. A. Garcia\textbf{$^{1}$, J. G. S. Duque$^{2}$, P. G. Pagliuso$^{1}$,
C. Rettori$^{1}$, Z. Fisk$^{3}$, and S. B. Oseroff$^{4}$.}}

\address{$^{1}$Instituto de Física {}``Gleb Wataghin'', C. P. 6165, UNICAMP,
Campinas-SP, 13083-970, Brasil.\\
$^{2}$Núcleo de Física, UFS, 49500-000, Itabaiana, SE, Brasil\\
$^{3}$University of California, Irvine, CA, 92697-4573, USA.\\
 $^{4}$San Diego State University, San Diego, California 92182, USA.}

\begin{abstract}
In this work we report electron spin resonance (ESR) measurements
in the semiconducting $Ce_{1-x}Gd_{x}Fe_{4}P_{12}$ ($x\approx0.001$)
filled skutterudite compounds. Investigation of the temperature ($T$)
dependence of the ESR spectra and relaxation process suggests, that
in the $T$-interval of $140-160$ K, the onset of a {}``weak''
metal-insulator (M-I) transition takes place due to the increasing
density of thermally activated carriers across the semiconducting
gap of $\approx$ 1500 K. In addition, the observed low-$T$ fine
and hyperfine structures start to collapse at $T\approx140$ K and
is completely absent for $T\gtrsim160$ K. We claim that the increasing
carrier density is able to trigger the rattling of the $Gd^{3+}$
ions which in turn is responsible, via a motional narrowing mechanism,
for the collapse of the ESR spectra.
\end{abstract}
\maketitle

\section{Introduction}

The filled skutterudite RT$_{4}$X$_{12}$ compounds, where R is a
rare-earth or actinide, T is a transition metal (Fe, Ru, Os) and X
is a pnictogen (P, As, Sb) have attracted great attention due to their
broad range of physical properties. They are of interest for those
seeking more efficient thermoelectric materials \cite{Snyder,Sales}
and also for those investigating the basic issues of strongly correlated
electron systems \cite{Goto,Bauer,Dilley}.

These compounds crystallize in the LaFe$_{4}$P$_{12}$ structure
with space group $Im3$ and local point symmetry T$_{h}$ for the
R ions. The R ions are guests in the oversized rigid cages constituted
by the (T$_{2}$X$_{3}$)$_{4}$ atoms \cite{Jeitschko}. The dynamics
of these guests R ions is believed to be of great importance in dampening
the thermal conductivity \cite{Lee,Herman} as observed in the filled
compounds of this family and it may also play an important role on
the appearance of heavy fermion behavior and superconductivity \cite{Goto,Yanagisawa}.

Electron spin resonance (ESR) is a sensitive and powerful microscopic
tool to provide information about crystal field (CF) effects, site
symmetries, valencies of paramagnetic ions, $g$-values, fine and
hyperfine parameters \cite{Bleaney}. In our recent work \cite{Garcia},
ESR was found to be a useful tool to probe the dynamics of the R ions.
The $T$-dependence of the guest R ions localization lead the ions
to experience slightly different symmetry environments causing a distribution
of $g$-values that was detected by our ESR experiment. In addition,
the remarkable reduction of the hyperfine parameters observed in the
ESR spectra was attributed to a motional narrowing mechanism \cite{Anderson}
caused by the rattling of the R ions. In our previous ESR experiments
on Yb$^{3+}$ impurities in Ce$_{1-x}$Yb$_{x}$Fe$_{4}$P$_{12}$
\cite{Garcia}, these two effects were analyzed and the coexistence
of two distinct sites were determined.

Ogita \emph{et al} \cite{Ogita}, performing Raman scattering experiments
on several metallic skutterudite compounds of RT$_{4}$X$_{12}$ (T
= Fe,Ru,Os; X = P,Sb), found resonant 2nd order phonon modes associated
with the vibration that change the bond length of R-X (stretching
mode). However, in the semiconducting CeFe$_{4}$P$_{12}$ the 2nd
order phonon modes were found to be non resonant. Based on these results
the authors conclude that there should be a strong coupling between
the R-X stretching modes and the conduction electrons. Thus, in CeFe$_{4}$P$_{12}$
a weak stretching mode-conduction electron coupling should be expected.

In this work, to further investigate these ideas, we have studied
the $T$-evolution of the ESR spectra of the Gd$^{3+}$ion in the
Ce$_{1-x}$Gd$_{x}$Fe$_{4}$P$_{12}$ ($x\approx0.001$) filled skutterudite
compounds.

\section{Experimental}

Single crystals of Ce$_{1-x}$Gd$_{x}$Fe$_{4}$P$_{12}$ ($x\lesssim0.001$)
were grown in Sn-flux as described in Ref. \cite{Meisner}. The cubic
structure ($Im3$) and phase purity were checked by x-ray powder diffraction.
The Gd concentrations were determined from the $H$ and $T$-dependence
of the magnetization, $M(H,T)$, measured in a Quantum Design SQUID
$dc$-magnetometer. The ESR experiments used crystals of $\sim$$2$x$2$x$2$
mm$^{3}$ of naturally grown crystallographic faces. The ESR spectra
were taken in Bruker X ($9.48$ GHz)-band spectrometer using appropriated
resonators coupled to a $T$-controller of a helium gas flux system
for $4.2$ K$\lesssim T\lesssim300$ K.

The low-$T$ ESR spectra show the full fine structure of Gd$^{3+}$and
its angular variation was studied at $T=4.2$ K, $T=20$ K and $T=300$
K with the applied field, $H$, in the (110) plane. The experimental
arrangement was set up in a way that $\theta=0$ corresponds to the
{[}001] direction. The complete results of this work will be the subject
of a more extended manuscript \cite{Garciaprep}. In this work we
will discuss measurements in the $T$-interval $110$ K$\lesssim T\lesssim160$
K for various directions of $H$. In this interval, the fine structure
collapses in one single broad line and its lineshape changes from
Lorentzian (insulator) to Dysonian (metallic). These features were
confirmed for 3 different batches.

\section{Results and Discussion}

Figure 1 presents the normalized low-$T$ ESR spectra ($T\approx4.2$
K) taken for $H$ along the {[}001] direction (figure 1a) and at $\theta\approx30^{\circ}$
from {[}001] in the (110) plane (figure 1b). The $^{155,157}$Gd$^{3+}$
(I = 3/2) isotopes hyperfine structure close to the central ($\frac{1}{2}\longleftrightarrow-\frac{1}{2}$)
transition is clearly observed in the inset of figure 1a, where a
zoom of the spectra around this transition is shown. The measured
hyperfine parameter is A = 5.5(2) Oe. This structure is also observed
(not so clearly) for the other fine structure components and it is
almost completely hidden on figure 1b, due to the overlap with the
other fine structure transitions. The inset of figure 1b shows that
some structure remains, distorting the lineshape of the observed single
line due to a crystal misorientation of $\approx1^{\circ}$.

\begin{figure}[t]
\includegraphics[clip,scale=0.2]{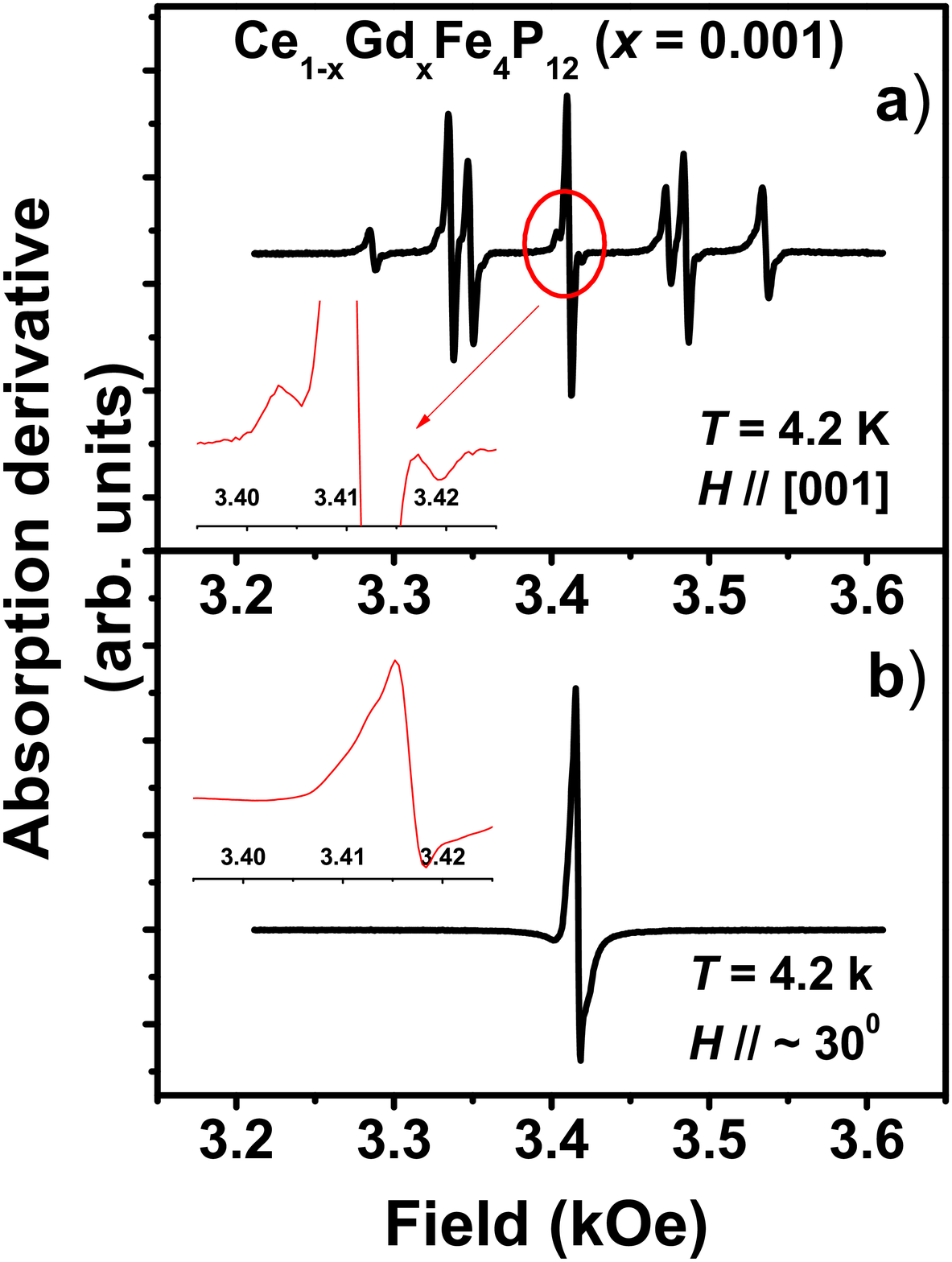} 

\caption{a) X-band ESR spectra ($T=4.2$ K) for $H$ along the {[}001] direction.
The inset shows a zoom of spectra that put in evidence the hyperfine
structure of the $^{157}$Gd$^{3+}$ (I = 3/2) isotope. b) X-band
ESR ($T=4.2$ K) for $H$ along $\theta=30^{o}$ from {[}001] in the
(110) plane. The inset presents details of the single line around
this angle.}

\label{figpaper1} 
\end{figure}

\begin{figure}[b]
\includegraphics[clip,scale=0.2]{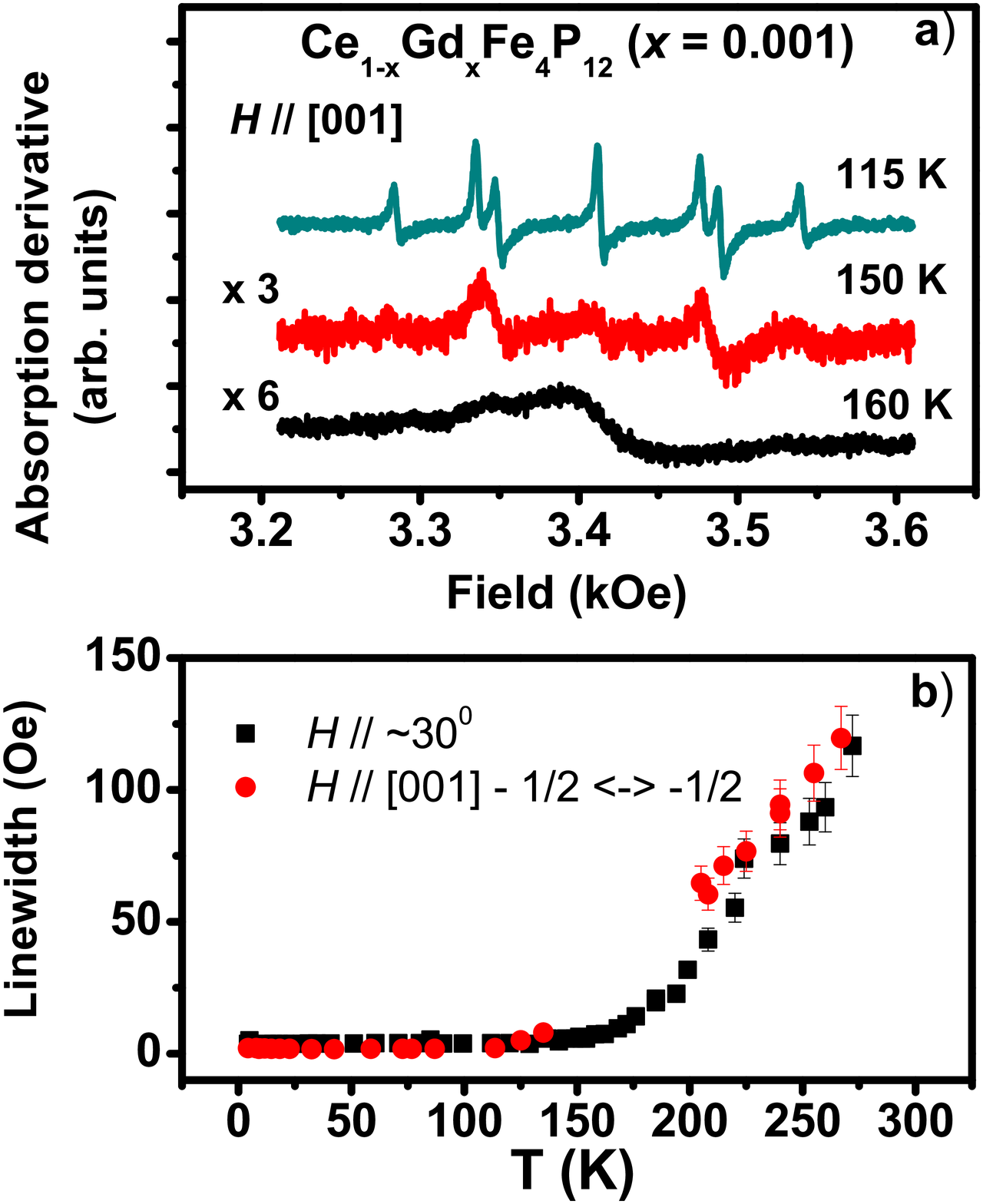} 

\caption{a) X-band $T$-dependence ESR spectra ($115\lesssim T\lesssim160$
K) for $H$ along the {[}001] direction. b) $T$-dependence of the
linewidth for $H$ along $\theta=30^{o}$ from {[}001] in the (110)
plane.}

\label{figpaper2} 
\end{figure}

Figure 2 shows the $T$-evolution of the above resonances in the $T$-interval
$140$ K $\lesssim T\lesssim160$ K. In figure 2a, for $H$ along
the {[}001] direction, the fine structure starts to collapse at $T\approx140$
K and is completely absent above $T\approx160$ K. The spectra turn
into a single broad line which broadens even further as $T$-increases.
In addition, the resonance lineshape throughout this $T$-interval
changes from Lorentzian (insulator) to Dysonian (metallic). Furthermore,
an isotropic (not shown) near linear $T$-dependence of the linewidth
of about 1.1 Oe/K is observed above $T\approx$ 200 K (see figure
2b). Notice that below $\simeq$ 140 K there is no $T$-dependence
of the linewidth. These results suggest that above $T\simeq$ 160
K the localized magnetic moment of the Gd$^{3+}$ ions may be relaxing
through an exchange interaction with the conduction electrons, i.e.
Korringa relaxation mechanism \cite{Korringa}.

Figure 3 presents the $T$-dependence of the hyperfine structure for
the ($\frac{1}{2}\longleftrightarrow-\frac{1}{2}$) transition. These
data show that the collapse of the hyperfine structure is observed
at $T\approx$ 10 K below the collapse of the fine structure and that
the ESR lineshape changes dramatically from Lorentzian (insulator)
to Dysonian (metallic).

The above results may be associated with the increasing metallic character
of CeFe$_{4}$P$_{12}$ due to the thermally activated carriers across
its semiconducting gap of $\approx$ 1500 K \cite{Sato}. Thus, we
suggest that the collapse of the fine and hyperfine structure, change
of the ESR lineshape and the Korringa-like relaxation are a consequence
of this increase in the carrier concentration. Following the Raman
experiments in metallic skutterudites \cite{Ogita}, we also suggest
that the increase in the metallic character of CeFe$_{4}$P$_{12}$
may activate the Gd-X stretching mode and, in turn, may trigger the
rattling of the Gd$^{3+}$ ions. The Gd$^{3+}$ rattling inside the
oversized (Fe$_{2}$P$_{3}$)$_{4}$ cage, via a motional narrowing
mechanism,\cite{Anderson} may be responsible for the collapse of
the fine and hyperfine structure into a single broad and isotropic
line \cite{Garcia}, whereas the Korringa-like relaxation for the
observed linear $T$-dependence of the linewidth \cite{Korringa}.

\begin{figure}[t]
\includegraphics[clip,scale=0.21]{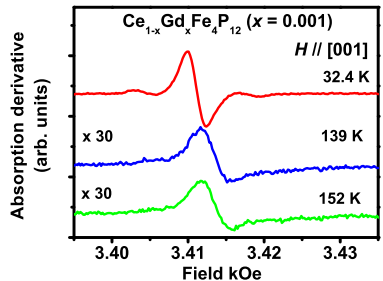} 

\caption{ESR spectra in the interval $32.4$ K$\lesssim T\lesssim155$ K, showing
the narrowing of the hyperfine structure for the $^{155,157}$Gd$^{3+}$
(I = 3/2) isotopes and the change in the lineshape of the ESR spectra.}

\label{figpaper3} 
\end{figure}

\section{Conclusions}

In general the $T$-dependence of the $dc$-resistivity in CeFe$_{4}$P$_{12}$
presents a semiconducting-like behavior. However, it is strongly sample
dependent and in some cases it shows a metallic behavior between $\sim$
50 and $\sim$ 200 K \cite{Sato}. Nevertheless, a common behavior
for all the samples is that above $\sim$ 200 K predominate the thermally
activated conductivity mechanism.

In this work we have presented experimental evidences that in the
$T$-interval of $140$ K$\lesssim T\lesssim160$ K there are three
dramatic changes in the ESR spectra: a) the collapse of the hyperfine
and fine structures; b) the ESR lineshape goes from Lorentzian (insulating
media) to Dysonian (metallic media) and c) the $T$-dependence of
the ESR linewidth changes from a $T$-independent to a nearly linear
$T$-dependence behavior which resembles the Korringa-like relaxation
process in a metallic host \cite{Korringa}. These results indicates
that around $\sim$ 160 K, and at the microwave frequency, there is
a clear change in the $ac$-conductivity of the material. We associate
this change to a \textquotedbl{}weak\textquotedbl{} metal-insulator
transition which could be detected by our highly sensitivity ESR experiment.

These observations, together with the Raman results on these compounds,
lead us to suggest that this \textquotedbl{}weak\textquotedbl{} metal-insulator
transition presumable triggers the rattling (stretching mode) of the
Gd$^{3+}$ ions in the oversized (Fe$_{2}$P$_{3}$)$_{4}$ cage and
that this rattling may be responsible for the collapse of the hyperfine
and fine structures via a motional narrowing mechanism \cite{Garcia,Anderson}.
The present study gives further insights for the subtle interaction
that in general exists between the localized vibration modes (Einstein
oscillators) of the R ions and the conduction electrons in the filled
skutterudite compounds. In particular, our work supports the idea
that some metallic character is always needed to set up the necessary
conditions for the rattling of the R ions in these materials.

\end{document}